\documentclass[showpacs,prd,nofootinbib,showkeys,twocolumn,10pt,aps,longbibliography]{revtex4-1}
\usepackage{hyperref} %create internal and external hyperlinks
\usepackage{amsmath}
\usepackage{graphicx}
\usepackage[caption=false]{subfig}
\usepackage{comment}
\usepackage{color}

\definecolor{darkgreen}{rgb}{0,.7,0}
\definecolor{linkblue}{rgb}{0.,0.,0.9333}

\hypersetup{
    pdffitwindow=true,      % page fit to window when opened
    pdfauthor={W. A. Horowitz},     % author
    pdfnewwindow=true,      % links to different pdfs in new window
    bookmarksnumbered=true,
    colorlinks=true,       % false: boxed links; true: colored links
    linkcolor=red,          % color of internal links
    citecolor=darkgreen,        % color of links to bibliography
    filecolor=magenta,      % color of file links
    urlcolor=cyan,           % color of external links
    menucolor=blue,
    baseurl=http://www.arxiv.org/abs/,
}

\newcommand{\fig}[1]{Fig.~\ref{fig:#1}}

\newcommand{\eq}[1]{Eq.~(\ref{eq:#1})}
\newcommand{\tab}[1]{Table~\ref{tab:#1}}
\newcommand{\infinity}{\infty}

\renewcommand{\Re}{\mathrm{Re}}
\renewcommand{\max}{\mathrm{max}}
\renewcommand{\min}{\mathrm{min}}

\begin{document}

\title {Fluctuating Heavy Quark Energy Loss in Strongly-Coupled Quark-Gluon Plasma}
\author{W.\ A.\ Horowitz}

\affiliation{Department of Physics, University of Cape Town, Private Bag X3, Rondebosch 7701, South Africa}
\email{wa.horowitz@uct.ac.za}

\begin{abstract}%
Results from an energy loss model that includes thermal fluctuations in the energy loss for heavy quarks in a strongly-coupled plasma are shown to be qualitatively consistent with single particle data from both RHIC and LHC.  The model used is the first to properly include the fluctuations in heavy quark energy loss as derived in string theory and that do not obey the usual fluctuation-dissipation relations.  These fluctuations are crucial for simultaneously describing both RHIC and LHC data; leading order drag results without fluctuations are falsified by current data.  
Including the fluctuations is non-trivial and relies on the Wong-Zakai theorem to fix the numerical Langevin implementation.  
The fluctuations lead to surprising results: $B$ meson anisotropy is similar to that for $D$ mesons at LHC, and the double ratio of $D$ to $B$ meson nuclear modification factors approaches unity more rapidly than even predictions from perturbative energy loss models.  It is clear that future work in improving heavy quark energy loss calculations in AdS/CFT to include all energy loss channels is needed and warranted.
\end{abstract}


\maketitle %\nonstopmode

\section{Introduction}
The goal of heavy ion collision research is to map out the many-body properties of quantum chromodynamics at extreme temperatures.  Simultaneously, this research provides insight into the confinement transition of QCD matter and yields information on the earliest moments of the Universe's existence: currently, no other experimental program can reach back as far in time \cite{LRP}.  

The fireballs created in central RHIC and LHC heavy ion collisions reach temperatures of a trillion degrees \cite{Hirano:2010je,Shen:2011eg,Qiu:2011hf,Gale:2012rq}; at this temperature the many-body dynamics of colored material is very different from that of normal nuclear matter \cite{Philipsen:2012nu}.  The best way of theoretically understanding this material is not yet known.  Simple questions such as ``What are the relevant degrees of freedom for the system?''\ do not yet have a clear answer.  At the most basic level, it is not clear how to reconcile experimental results that, na\"ively, simultaneously suggest a strongly-coupled plasma that evolves hydrodynamically and a weakly-coupled gas of slightly modified quarks and gluons that yields a plasma relatively transparent to hard probes.  

In order to make progress as a field it is necessary to attempt to effect such a reconciliation of ideas of the nature of the plasma into a consistent theoretical picture that yields quantitatively consistent comparisons with experimental data.  Focusing on two major avenues of investigations, low transverse momentum (low-$p_T$) observables that are described well by near-ideal relativistic hydrodynamics switched on very soon after the initial collision overlap \cite{Hirano:2010je,Shen:2011eg,Qiu:2011hf,Gale:2012rq} and high-$p_T$ observables associated with hard production described by perturbative quantum chromodynamics (pQCD) \cite{Majumder:2010qh,Horowitz:2012cf,Djordjevic:2014tka}, the low-$p_T$ observables appear readily understood within a strong-coupling paradigm for the plasma in which calculations are performed using the AdS/CFT correspondence \cite{Gubser:1996de,Kovtun:2004de,CasalderreySolana:2011us} while the high-$p_T$ observables appear readily understood within a weak-coupling paradigm for the plasma in which calculations are performed using pQCD \cite{Majumder:2010qh}.  

Specifically, leading order derivations exploiting the AdS/CFT correspondence found that quark-gluon plasma (QGP)-like systems have thermodynamic properties such as entropy density similar to those seen from the lattice \cite{Gubser:1996de,Philipsen:2012nu}; rapidly hydrodynamize after a heavy ion collision-like event \cite{Chesler:2010bi}, i.e.\ nearly ideal relativistic hydrodynamics approximates well the dynamics of the system soon after a heavy ion collision as is inferred from data; and have a very small shear viscosity-to-entropy ratio $\eta/s\,\sim\,0.1$ in natural units \cite{Gubser:1996de,Kovtun:2004de}, again as inferred from hydrodynamics studies of data \cite{Hirano:2010je,Shen:2011eg,Qiu:2011hf,Gale:2012rq}.  At the same time, leading order derivations based on pQCD show differences to thermodynamic properties \cite{Cheng:2007jq}, do not hydrodynamize rapidly \cite{Attems:2012js}, and have a viscosity-to-entropy ratio $\eta/s\,\sim\,1$ \cite{Danielewicz:1984ww} at least an order of magnitude larger than that inferred from hydrodynamics.

On the other hand, energy loss models for light and heavy flavored particles based on calculations using leading order pQCD show broad qualitative agreement with data from both RHIC and LHC \cite{Majumder:2010qh,Horowitz:2012cf,Djordjevic:2014tka}.  At the same time, early light flavor \cite{Horowitz:2011cv} and leading order heavy flavor \cite{Horowitz:2011wm} energy loss models using AdS/CFT found or suggested a massive oversuppression of high-$p_T$ leading particles compared to data.

Much work has been done to improve perturbative field theoretic treatments of soft, or low-$p_T$, observables.  For example there are high-order thermal field theory calculations of thermodynamic properties of the QGP \cite{Haque:2014rua} and recent work suggested that pQCD can yield rapid isotropization times of $0.2-1.0$ fm \cite{Kurkela:2014tea}.  The small viscosity-to-entropy ratio inferred from hydrodynamics remains a challenge for perturbative calculations; no pQCD calculation has yet yielded $\eta/s\,\sim\,0.1$.

An alternative research avenue is to improve the AdS/CFT treatment of high-$p_T$ observables.  There are hybrid calculations in which, for instance, AdS/CFT is used to model the medium only \cite{Gubser:2006nz,CasalderreySolana:2007qw,D'Eramo:2010ak} or only part of the energy loss \cite{Casalderrey-Solana:2014bpa}, although in the latter case the coupling is taken to be quite small, $\lambda\,\sim\,10^{-3}-10^{-4}$.  Other work has attempted to apply AdS/CFT to the entire energy loss problem.  A recent success \cite{Morad:2014xla} found that when the strong-coupling jet prescription is improved and the in-medium energy loss is renormalized, a reasonable value of $\lambda \, = \, 5.5$ yielded a jet nuclear modification factor $R_{AA}^{jet}(p_T)$ quantitatively consistent with preliminary CMS data \cite{CMS:2012rba}.

By allowing the variation of the intrinsic mass of the probe, heavy quark observables provide an extremely valuable additional test of energy loss calculations and, therefore, the properties of quark-gluon plasma.  It has been very fruitful for the energy loss community to require of energy loss models a simultaneous description of both light and heavy flavor observables.  For example, \cite{Djordjevic:2005db} showed that pQCD-based energy loss models that only include the radiative energy loss channel could not simultaneously describe both suppression of particles associated with both light and heavy flavors at RHIC.  From the strong-coupling side of calculations, heavy quark observables are especially useful as the theory of heavy quark energy loss is both more mature and under better control than that for light flavors \cite{Gubser:2006bz,Herzog:2006gh,CasalderreySolana:2006rq,Chesler:2008uy,Ficnar:2012np,Morad:2014xla}.  

The work of this paper is to improve the energy loss modelling of heavy quarks under the assumption of a strongly-coupled medium strongly-coupled to the heavy quark.  Previous calculations that used only the leading order strongly-coupled heavy quark energy loss, often referred to as heavy quark drag, compared favorably to the measured nuclear modification factor $R_{AA}(p_T)$ of electrons from heavy flavor decay at RHIC \cite{Horowitz:2010dm} but generally oversuppressed $R_{AA}(p_T)$ for $D$ mesons at LHC by a factor of $\sim\,5$ \cite{Horowitz:2011wm}.  Another early comparison to RHIC data included fluctuations in the energy loss as dictated by the fluctuation-dissipation theorem \cite{Akamatsu:2008ge}.  These results were consistent with data from RHIC, but were never compared to data from LHC.  A major deficiency of the approach of \cite{Akamatsu:2008ge} is that the string theoretic derivation of the thermal momentum fluctuations for heavy quarks \cite{Gubser:2006nz} demonstrated that the momentum diffusion is not related to the momentum drag according to the fluctuation-dissipation theorem.  Even worse, the fluctuations are non-Markovian; the fluctuations are due to colored noise, which means that there are non-trivial temporal correlations between the momentum kicks.

In this paper the correct fluctuation spectrum is used and the non-Markovian nature of the momentum fluctuations is mitigated; it will be shown that by doing so one can qualitatively describe the data related to single heavy flavor observables at both RHIC and LHC simultaneously.   

\section{Langevin Energy Loss}
It was shown in \cite{Gubser:2006nz} that the three-momentum $p^i$ of an on-shell heavy quark moving at constant velocity in a thermal bath evolves as:
\begin{align}
	\label{eq:langevin}
	\frac{dp_i}{dt} = -\mu \, p_i + F_i^L+F_i^T,
\end{align}
where the drag coefficient \cite{Gubser:2006bz,Herzog:2006gh,CasalderreySolana:2006rq} of a heavy quark of mass $M_Q$ in a plasma of temperature $T$ with 't Hooft coupling $\lambda$ is
\begin{align}
	\label{eq:drag}
	\mu = \frac{\pi\sqrt{\lambda}T^2}{2M_Q}
\end{align}
and the fluctuating momentum kicks are correlated as
\begin{align}
	\langle F_i^L(t_1) F_j^L(t_1) \rangle & = \kappa_L \hat{p}_i \hat{p}_j g(t_2-t_1) \\
	\langle F_i^T(t_1) F_j^T(t_1) \rangle & = \kappa_T (\delta_{ij}-\hat{p}_i \hat{p}_j) g(t_2-t_1),
\end{align}
where $\hat{p}_i \, = \, p_i / |\vec{p}|$,
\begin{align}
	\kappa_T & = \pi \sqrt{\lambda} T^3 \gamma^{1/2} \\
	\label{eq:kappaL}
	\kappa_L & = \gamma^2\kappa_T,
\end{align}
and $g$ is a function known only numerically.

Since the fluctuations increase in magnitude with heavy quark velocity---in fact increase very, very quickly in the longitudinal direction (like $\gamma^{5/2}$), which is the most important direction for calculations of suppression observables---one may naturally ask: at what speed should one expect the fluctuations to play a nontrivial role in heavy quark energy loss?  This speed limit may be estimated by requiring that the momentum picked up via fluctuations over the time scale set by the drag coefficient be small in comparison to the total momentum of the heavy quark.  One finds then an upper limit on the speed of the heavy quark given by
\begin{equation}
	\label{eq:flucspeedlimit}
	\gamma \lesssim \gamma_{crit}^{fluc} = \frac{M_Q^2}{4T^2}.
\end{equation}

There is a well-known speed limit for the heavy quark drag setup in which the derivation of \eq{langevin} was performed.  A finite mass heavy quark is represented by a string connected to a D7 brane that partially fills the fifth dimension \cite{Karch:2002sh}.  The mass of the heavy quark is related to the depth that the D7 brane fills the fifth dimension; the greater the depth, the lighter the quark.  In a space with a black hole---dual to a plasma at finite temperature \cite{Witten:1998zw}---the local speed of light decreases with the depth of the fifth dimension.  Thus in this setup the endpoint of a string dual to a finite mass heavy quark that is connected to the bottom of the D7 brane has as a natural speed limit the local speed of light \cite{Gubser:2006nz}; a heavy quark in the field may move faster than this speed, but then the setup is necessarily incorrect for this physical situation.  Equivalently, the speed limit can be derived from self-consistency: in the setup in which the drag formulae are derived, the heavy quark moves at a constant velocity.  In order for the constant velocity to be maintained, one must provide power to the heavy quark to balance the momentum lost.  If that power is provided by a gauge field, there is a maximum field strength before which Schwinger pair production begins; that maximum field strength limits the velocity with which the quark can be pulled through the plasma at a constant velocity \cite{CasalderreySolana:2007qw}.  Both lines of reasoning lead to exactly the same speed limit.  This speed limit on the heavy quark setup leads to the requirement that
\begin{equation}
	\label{eq:speedlimit}
	\gamma < \gamma_{crit}^{sl} = \big( 1 + \frac{2M_Q}{\sqrt{\lambda}T}\big)^2 \sim \frac{4M_Q^2}{\lambda T^2}.
\end{equation}

One may easily see that \emph{parametrically} the critical speed is restricted most by the speed limit due to the requirement of constant heavy quark velocity, not due to the momentum kicks picked up from the medium; the former goes as $\lambda^{-1}$ while the latter goes as $\lambda^0$.  However, in the case of finite $\lambda\sim\mathcal{O}(10)$ one finds that the critical speed due to momentum fluctuations is actually smaller than that of the speed limit.  It is therefore important to include these fluctuations in a phenomenological heavy quark energy loss model based on AdS/CFT that is compared to data, the goal of this work.

\begin{figure}[!htbp]
    \subfloat[][]{
    \includegraphics[width=\columnwidth]{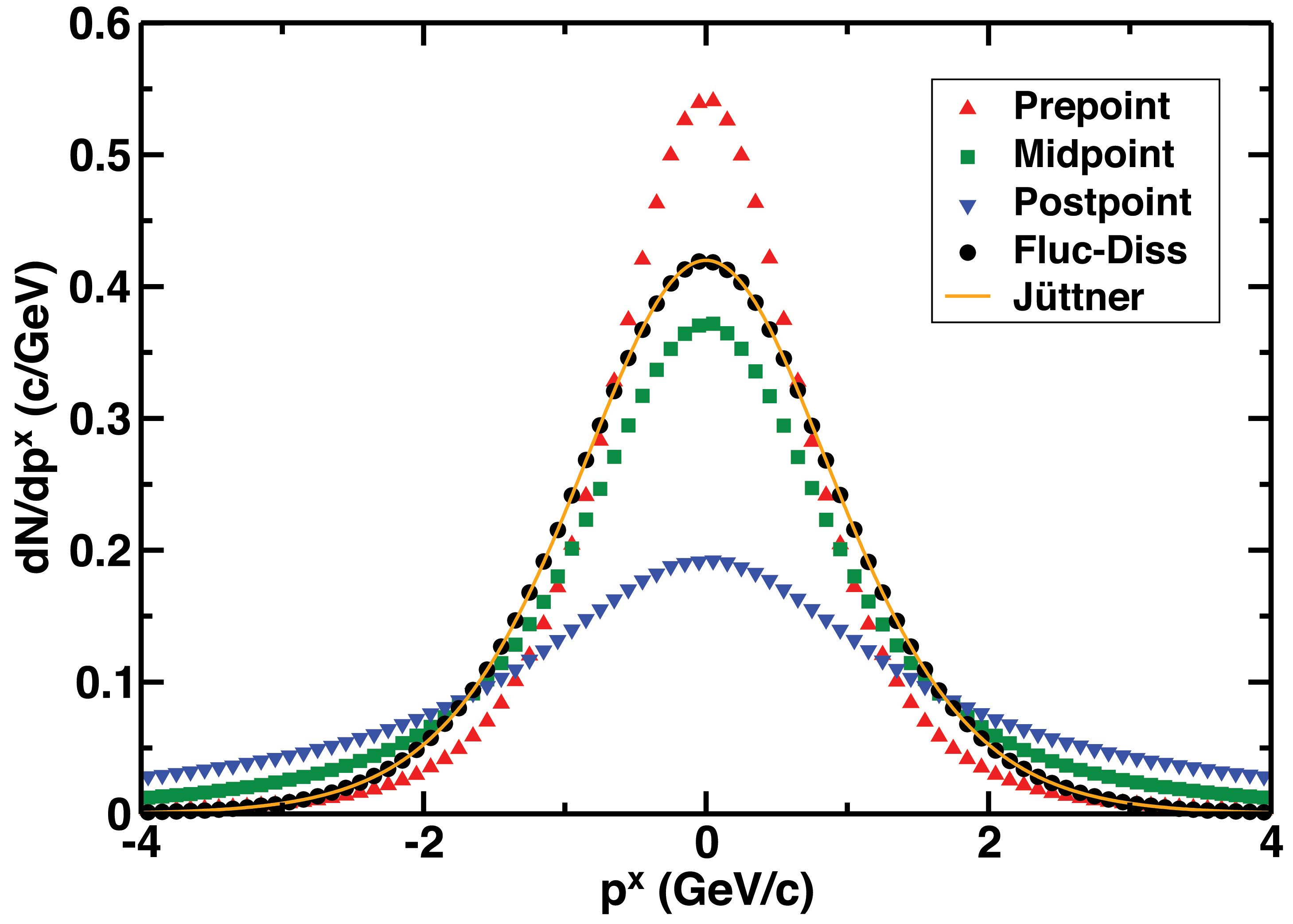}
    \label{subfig:dNdpx}
    }\\
 	\subfloat[][]{
 	\includegraphics[width=\columnwidth]{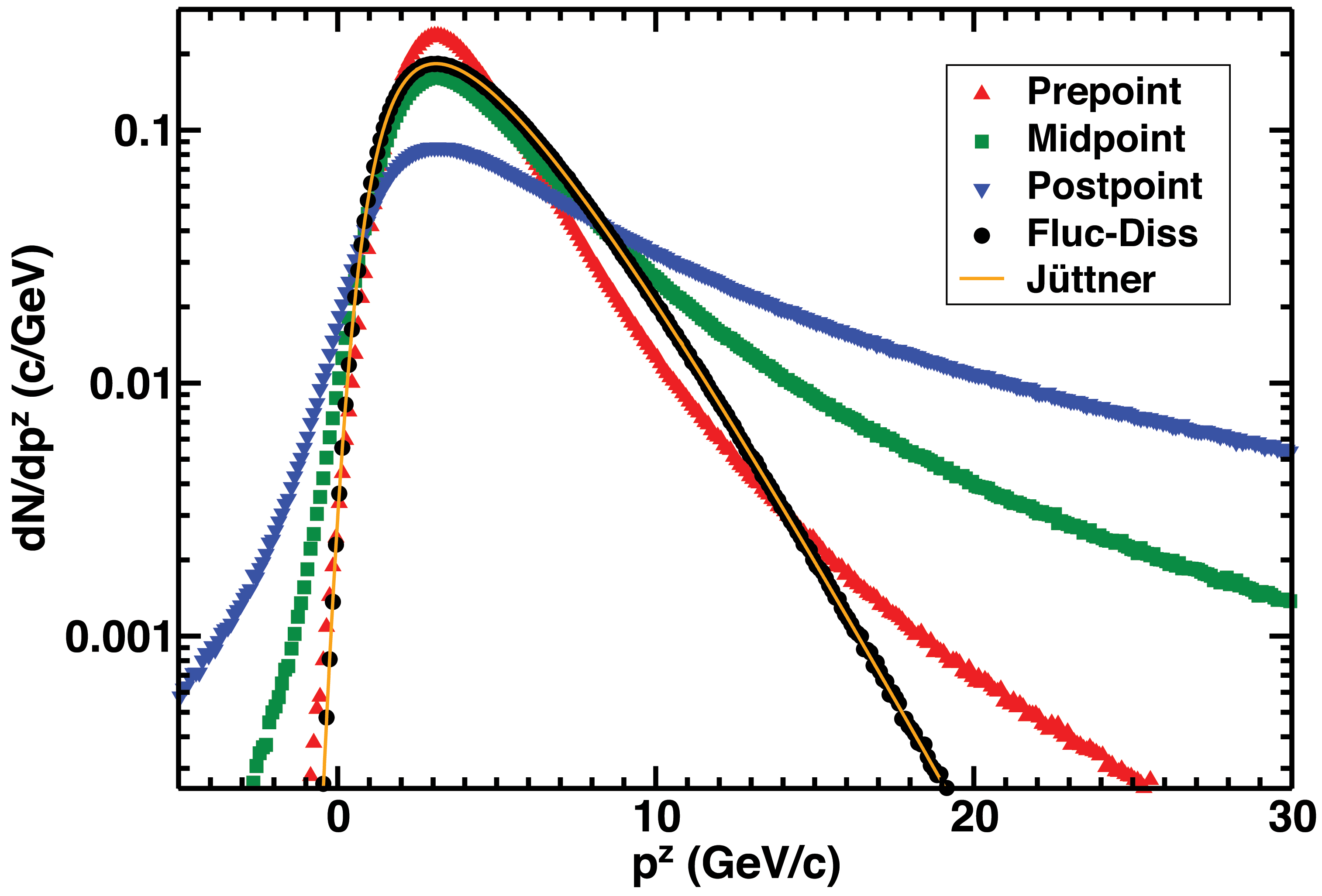}
 	\label{subfig:dNdpz}
 	}
     \caption{\label{fig:thermal}(\protect\subref*{subfig:dNdpx}) The $p^x$-differential distribution of $m_c \, = \, 1.5$ GeV$/$c$^2$ charm quarks strongly-coupled to a strongly-coupled plasma in three spatial dimensions at $T\, = \, 400$ MeV which is moving at $0.9\,$c in the $z$ direction.  The effect of the thermal drag and fluctuations on the heavy quarks from strong coupling is found from implementing \protect\eq{langevin} using the It\^o (pre-point), Stratonovich (mid-point), and H\"anggi-Klimontovich (post-point) stochastic integrals.  An implementation for which the fluctuation-dissipation theorem holds is also shown and compared to the result expected, the J\"uttner distribution. (\protect\subref*{subfig:dNdpz}) The same as (\protect\subref*{subfig:dNdpx}) but for the $p^z$-differential distribution.}	
\end{figure}

The implementation of the fluctuations in \eq{langevin} are nontrivial for two reasons: first, the noise is \emph{colored} and second, the fluctuations are \emph{multiplicative}.  With respect to the former complication, if $g$ was a Dirac delta function, the momentum fluctuations in \eq{langevin} would correspond to white noise; i.e.\ one momentum kick is not correlated in time with any other.  However, it was shown in \cite{Gubser:2006nz} that the noise is actually colored, in which case $g$ is some nontrivial function of characteristic width $t_{corr} \sim \sqrt{\gamma}/T$.  With respect to the latter complication, multiplicative noise has correlations that depend on the dynamical variable, in this case $p_i$.  Unlike usual Riemann calculus, stochastic Ito integration is sensitive to the precise location at which the integrand is evaluated within a time step; different choices of location within the time step yield different results for the integral \cite{Evans,Gardiner,vanKampen,Kloeden}.  Note that these different results for the integral are truly fundamental and not due to finite size time steps.  Unfortunately, the work of \cite{Gubser:2006nz} did not specify at which place within the time step the derived fluctuations should be applied.  \fig{thermal} shows the significant differences in the momentum distributions of charm quarks in a thermal plasma of $T \, = \, 400$ MeV which is moving at $v \, = \, 0.9\,$c in the $z$ direction for three different but common choices made for the location within the time step to apply the momentum fluctuation.  In particular the results for the It\^o (pre-point, for which the integrand and the fluctuation are evaluated at the earliest time within a time step), Stratonovich (mid-point, for which the integrand and the fluctuation are evaluated at the middle of a time step), and H\"anggi-Klimontovich (post-point, for which the integrand and the fluctuation are evaluated at the latest time within a time step) rules are shown.  The differences between rules are especially pronounced in the tails of the distribution (c.f.\ \fig{thermal} (\subref*{subfig:dNdpz})), which tends to be the most important for the purposes of energy loss modeling.  Note that none of the distributions produced by the different integration prescriptions for the momentum fluctuations given by AdS/CFT are exactly thermal.  The deviation away from the thermal distribution can be attributed at least in part to the modification of the usual dispersion relation of the heavy quark due to its coupling to the strongly-coupled thermal medium \cite{Herzog:2006gh}.  In \fig{thermal} the relativistic Maxwell-Boltzmann, or J\"uttner, distribution is shown and compared to a distribution produced by a Langevin evaluation whose diffusion coefficient is related to its drag via the relativistic fluctuation-dissipation relation \cite{He:2013zua}.  This comparison between the analytic and numerical results is a non-trivial cross check of both the analytics and numerics: the distribution along the direction of plasma fluid flow, \fig{thermal} (\subref*{subfig:dNdpz}), is not a trivial boost of the distribution along a direction for which the plasma is at rest, \fig{thermal} (\subref*{subfig:dNdpz}); see Appendix \ref{juttner}.  

Fortunately, both the colored and multiplicative complications of the noise can be handled.  As for the first complication, in every physical situation, all noise is actually colored (for example the air molecule that just collided with a floating pollen grain cannot immediately collide with the pollen grain again); the issue is whether or not the time scale over which the momentum kicks are correlated is small compared to the other relevant time scale(s) in the problem.  For heavy quark energy loss as computed at strong coupling in AdS/CFT, the only other relevant time scale is determined by the drag coefficient $\mu$ from \eq{langevin}.  One may check that
\begin{equation}
	\label{eq:autocorr}
	t_{corr}\mu \sim \frac{1}{2}\sqrt{\lambda}\sqrt{\gamma}\frac{T}{M_Q} < 1
\end{equation}
so long as $\gamma \, < \, \gamma_{crit}^{sl}$ from \eq{speedlimit}.  Thus we are able to approximate the colored noise as white noise so long as the heavy quark is moving slower than the speed limit imposed by the setup of the derivation. As for the second complication, there is a theorem due to Wong and Zakai \cite{WongZakai} that proves that in the limit that the autocorrelation time for noise goes to zero, the correct stochastic integral is the Stratonovich one, which is to say that the integrand and noise terms are evaluated at the midpoint of the time step.  As just noted, for the specific calculation of interest in this work, the autocorrelation time is small so long as the heavy quark moves more slowly than the speed limit $\gamma_{crit}$, \eq{speedlimit}.  

In the energy loss model used in this paper, the stochastic differential equation \eq{langevin} was solved numerically using the Euler-Maruyama scheme \cite{Kloeden}.  Despite its simplicity, the Euler-Maruyama scheme converges strongly, and the scheme was sufficient for the purposes in this paper.  (Higher order strongly convergent schemes exist, e.g.\ Milstein, but are highly non-trivial to implement for motion in more than one dimension.)  In any implementation of a stochastic integral, whether It\^o, H\"anggi-Klimontovich, or anything in-between, one may actually evaluate the integral in any other choice with the addition (or subtraction) of the appropriate terms \cite{Kloeden,Gardiner}.  In particular an It\^o (pre-point) stochastic differential equation
\begin{equation}
	dp^i = a^i\big( t,\, p^k(t) \big)dt + b^{ij}\big( t,\, p^k(t) \big)dW_j
\end{equation}
with Wiener process $dW^j$ is equivalent to the Stratonovich stochastic differential equation of the form
\begin{equation}
	dp^i = \underline{a}^i\big( t,\, p^k(t) \big)dt + b^{ij}\big( t,\, p^k(t) \big)\circ dW_j,
\end{equation}
where
\begin{multline}
	\label{eq:itotostr}
	a^i\big( t,\, p^k(t) \big) = \underline{a}^i\big( t,\, p^k(t) \big) \\ + \frac{1}{2}\sum_\ell \sum_m b^{\ell m}\big(t,\, p^k(t)\big)\frac{\partial b^{im}}{\partial p^\ell}\big(t,\, p^k(t)\big).
\end{multline}
For ease of numerics, all results shown in this work had the integrand and noise terms evaluated at the earliest moment in the time step (It\^o, or, pre-point).  Thus in the local fluid rest frame the momentum components of a heavy quark at the next time step, $p'^i_{n+1}$, were found from the momenta at the current time step, $p'^i_n$, by interpreting \eq{langevin} as a Stratonovich stochastic differential equation, modified by \eq{itotostr} to be implemented as an It\^o SDE in the Euler-Maruyama scheme with the result that
\begin{multline}
	p'^i_{n+1} = \Big(1 - \mu \, dt' + \frac{1}{2} \kappa \, dt' \big( \, \frac{5\gamma^{5/2}}{4E'^2} \\ + \frac{(d-1) \, \gamma^{1/2}}{(\gamma^2+1) \, m^2} \, \big)  \Big) p'^i_n + C^{ij} dW_j,
\end{multline}
where $\mu$ is given by \eq{drag}; $dt'$ is the time step $dt$ boosted into the local rest frame of the fluid; $dt' \, = \, dt/\gamma$; $\kappa \, = \, \pi\sqrt{\lambda}T^3$, where $T$ is the temperature of the fluid in its local rest frame; $d$ is the number of spatial dimensions in the calculation (in this case we propagate the heavy quarks through backgrounds generated by VISHNU \cite{Shen:2011eg,Qiu:2011hf}, which is a $2+1D$ hydrodynamics code); the $dW_j$ are the uncorrelated, Gaussian Wiener kicks with mean 0 and standard deviation 1; and
\begin{align}
	C^{ij} & = \sqrt{dt'\,\kappa}\gamma^{1/4}\left( \frac{p'^ip'^j}{(\gamma^2+1) \, m^2} + \delta^{ij}\right).
\end{align}
In a calculation that follows the trajectory of the heavy quark from production onwards, at each time step the code boosts into the local rest frame of the fluid, evaluates the change in momentum, boosts back to the lab frame, and propagates the heavy quark in coordinate space according to
\begin{align}
	x^i_{n+1} = \frac{p^i_{n+1}}{E_{n+1}}dt.
\end{align}
$dt$ was taken to be $1/150\times\mu_{max}$, where $\mu_{max}$ is the drag coefficient at the center of the fireball at the thermalization time, the largest drag coefficient for any individual collision.  This value of $dt$ was found through trial and error to consistently yield stable static thermal distributions such as those shown in \fig{thermal}.  Variations of $dt$ by a factor of 2 in both directions yielded unchanged results in trials.  The code was written in C++%\footnote{All code necessary to run the energy loss model can be found at \url{http://www.phy.uct.ac.za/people/horowitz}}
, compiled using Clang, and run on a Mac laptop.

\section{Energy Loss Model}
The results presented here are from a fully Monte-Carlo energy loss model; only the fragmentation from heavy quarks into their decay products was handled through (numerical) integration of analytic expressions.  In particular, the production spectrum for heavy quarks (from FONLL \cite{Cacciari:1998it,Cacciari:2001td,Cacciari:2012ny}), the spatial distribution of their production points (weighed by the Glauber binary distribution \cite{Miller:2007ri}), and their initial direction of propagation (assumed uniform) were randomly sampled.  Propagation was through backgrounds generated by the VISHNU 2+1D hydrodynamics code \cite{Shen:2011eg,Qiu:2011hf}, with momenta updated according to \eq{langevin}.  (Pseudo-) random number generation was performed using the \texttt{Ran} routine from Numerical Recipes \cite{numericalrecipes}.  Seed generation came from the routine described in \cite{Katzgraber}, which is appropriate for running multiple processes at or nearly at identical times on multiple processors\footnote{Random number generators are often seeded by the run time of the code; when multiple instances of the code are created at the same time, runs can have additional, unintended correlation from using the same seed.}.

The production spectrum and fragmentation functions of the heavy quarks from FONLL were used \cite{Cacciari:1998it,Cacciari:2001td,Cacciari:2012ny}.  The spectra were generated by the FONLL web page \cite{FONLLwebpage} for the relevant center of mass and rapidity ranges for the various experiments.  The fragmentation functions from heavy quarks to heavy mesons were taken from \cite{Cacciari:2012ny}.  For measurements of unspecified $D$ mesons (and for electrons from $D$ mesons), the FONLL approximation of $D$ mesons being comprised of 70\% $D^0$ and 30\% $D^+$ mesons was used \cite{FONLLwebpage}.  The semi-leptonic decay of heavy mesons to electrons is not given explicitly in any FONLL paper.  The probability $P(E)$ of a heavy meson decaying to an electron of energy $E$ in the rest frame of the heavy meson used in \cite{Cacciari:2005rk} was provided \cite{VogtCacc}, but detailed instructions for implementation were not.  One may calculate the distribution of electrons in the lab frame from the semi-leptonic decay of a heavy meson; the details are given in Appendix \ref{ffs}.  In particular, Eqs.\ (\ref{eq:elecuse}) and (\ref{eq:pofd}) -- (\ref{eq:pofjpsi}) are used throughout the rest of this work for the fragmentation of heavy mesons to electrons and $B$ mesons to $J/\psi$ mesons.

\begin{figure}[!htbp]
	\centering
	\includegraphics[width=\columnwidth]{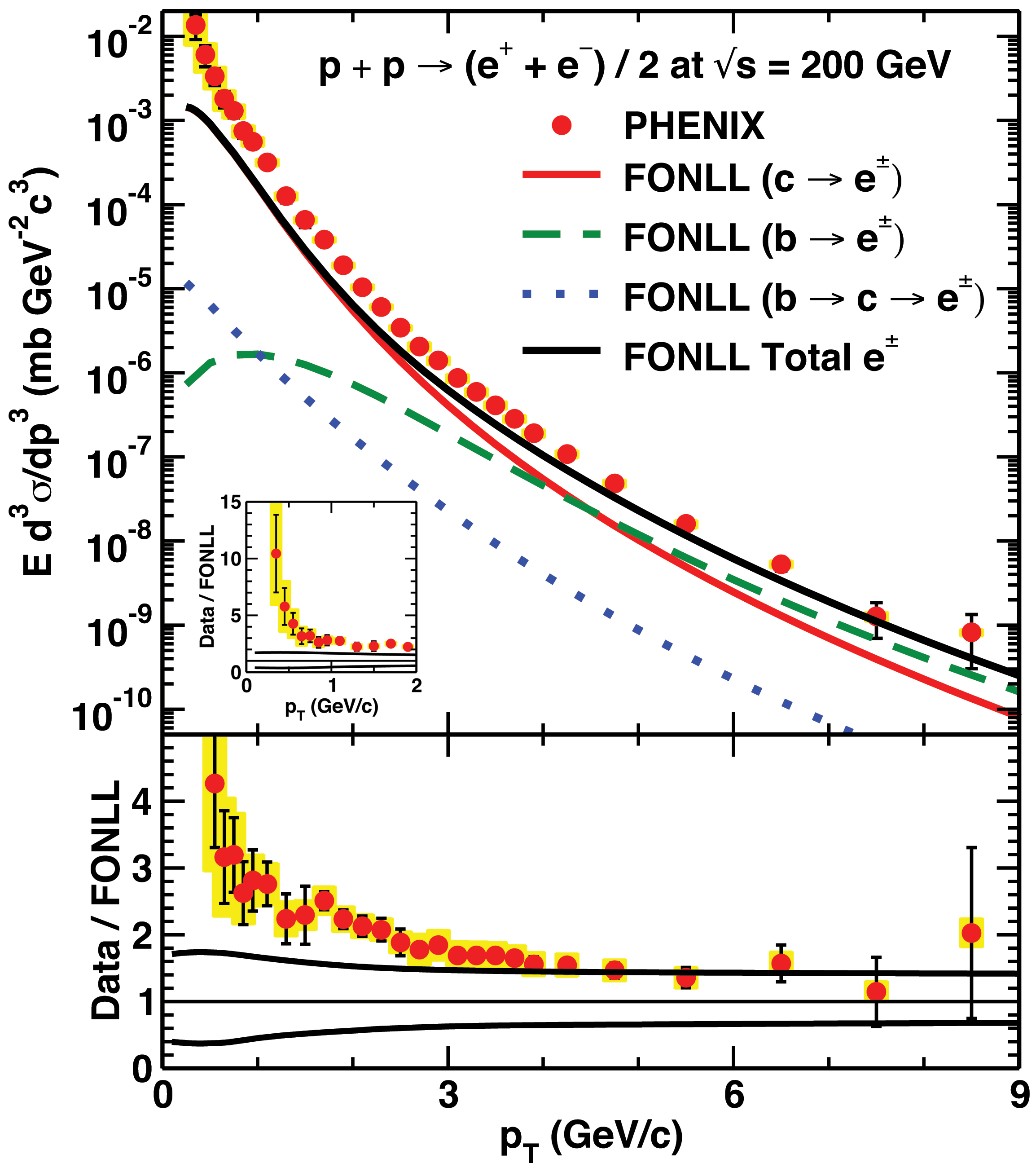}
	\caption{\label{fig:electrons} (Top) A comparison of FONLL predictions \protect\cite{Cacciari:1998it,Cacciari:2001td,Cacciari:2012ny} for the electron spectrum from open heavy flavor production in $p+p$ collisions at $\sqrt{s} \, = \, 200$ GeV to that measured by PHENIX \protect\cite{Adare:2010de}. (Bottom) Ratio of the PHENIX data to FONLL predictions; the solid black lines represent the uncertainty in the FONLL predictions due to scale variations. (Inset) Zoom in on the low-$p_T$ part of the ratio of data to FONLL predictions showing the full range of disagreement.}
\end{figure}

The production spectrum of electrons originating from open heavy flavor in $p+p$ collisions at $\sqrt{s} \, = \, 200$ GeV from FONLL is compared to that measured by PHENIX \cite{Adare:2010de} at RHIC in \fig{electrons}.  One can see that at ultrarelativistic momenta, the FONLL predictions are consistent with the data within the experimental and theoretical uncertainties, which amount to approximately a factor of 2.  At lower momenta the FONLL predictions deviate more, rising to a discrepancy of an order of magnitude for the smallest measured momentum bin of $p_T \, = \, 0.25$ GeV/c.  For electrons it appears that the production and/or the fragmentation description is not reliable below $p_T^e \, \lesssim \, 3$ GeV/c.  (It seems likely that the issue is with the fragmentation to electrons, with the discrepancy made much worse by the experimentally restricted rapidity range.)

When computing the energy loss and fluctuations, one needs a mapping from the QCD parameters to the $\mathcal{N} \, = \, 4$ SYM parameters.  Results in this paper are shown using two prescriptions, which are denoted in the plots ``$\alpha_s \, = \, 0.3$'' and ``$\lambda \, = \, 5.5$,'' in order to explore a reasonable breadth of possibilities (one hopes that a calculation in a theory dual to QCD would lie somewhere between these two prescriptions).  For the former, the 't Hooft coupling is taken to be $\lambda \, = \, 4\pi\alpha_s N_c \, = \, 4\pi\times0.3\times3$ and $T_{QCD} \, = \, T_{SYM}$.  For the latter, the prescription of \cite{Gubser:2006nz} is followed, in which $\lambda \, = \, 5.5$ and $T_{SYM} \, = \, T_{QCD}/3^{1/4}$.  The masses of the quarks were taken to be $m_c \, = \, 1.5$ GeV$/$c$^2$ and $m_b \, = \, 4.75$ GeV$/$c$^2$.

\section{Results}

\begin{figure}[!htbp]
	\subfloat[][]{
		\includegraphics[width=\columnwidth]{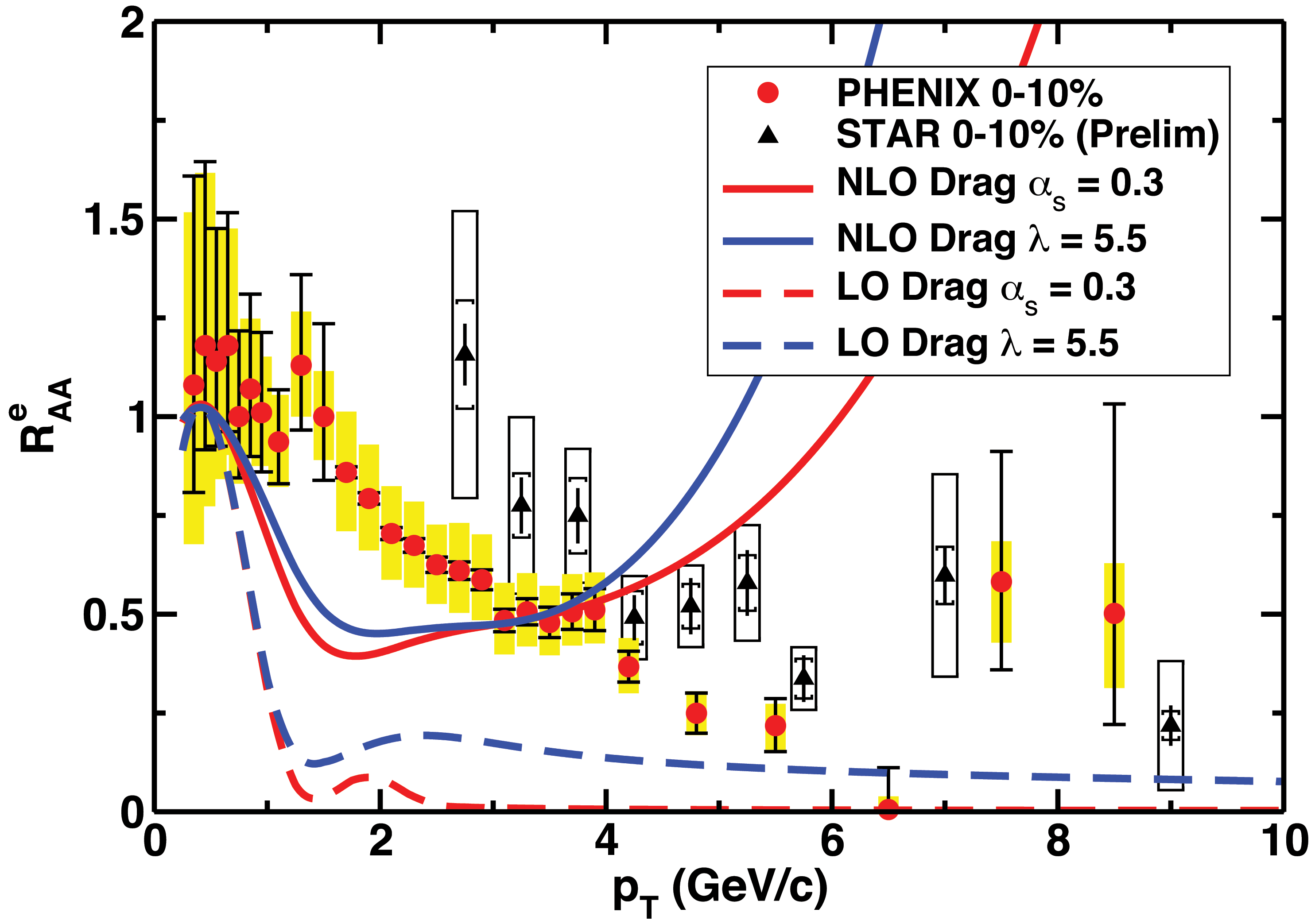}
		\label{subfig:RHICe}
	}\\[-.02in]
	\subfloat[][]{
		\includegraphics[width=\columnwidth]{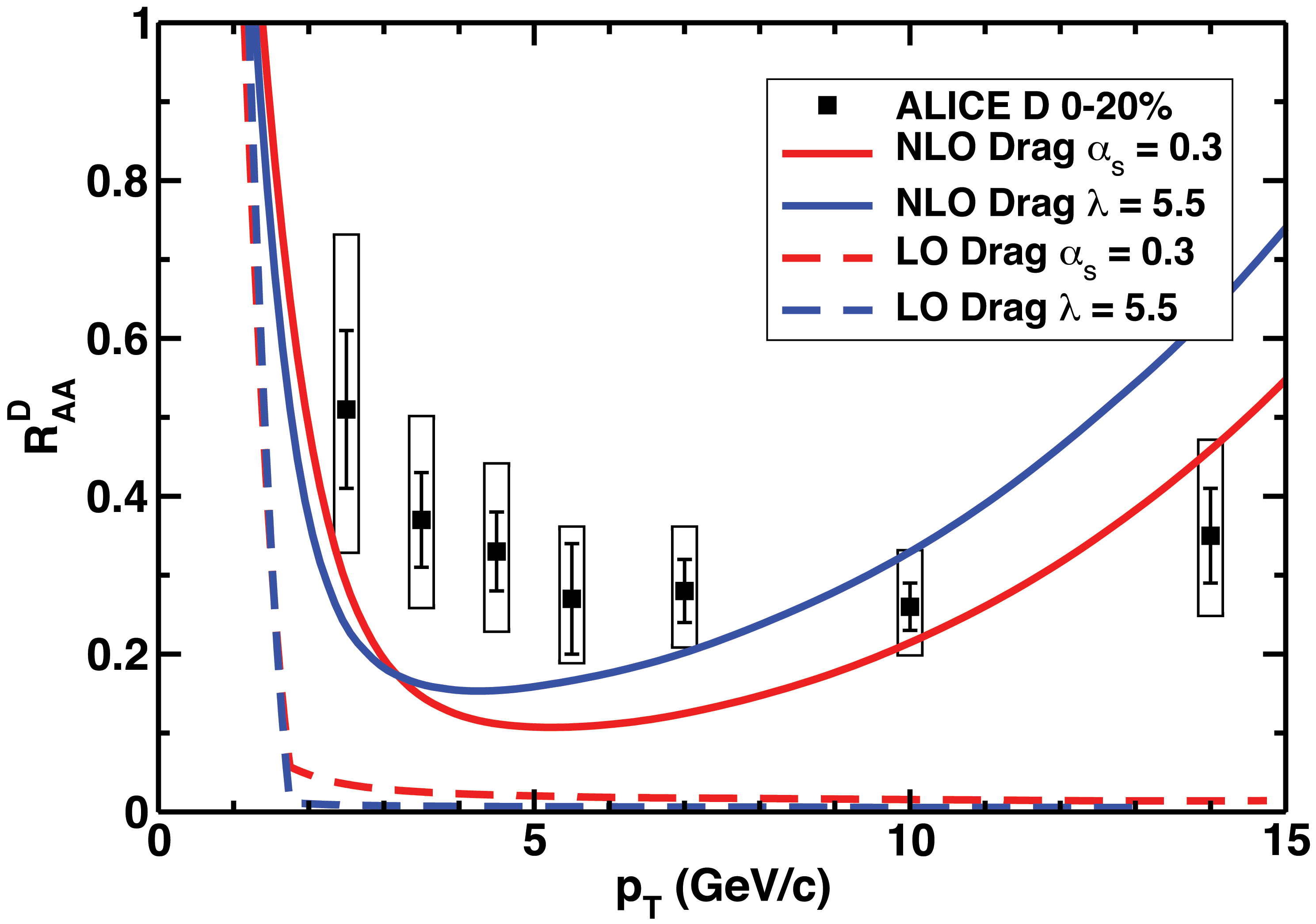}
		\label{subfig:LHCD}
	}\\[-.01in]
	\subfloat[][]{
		\includegraphics[width=\columnwidth]{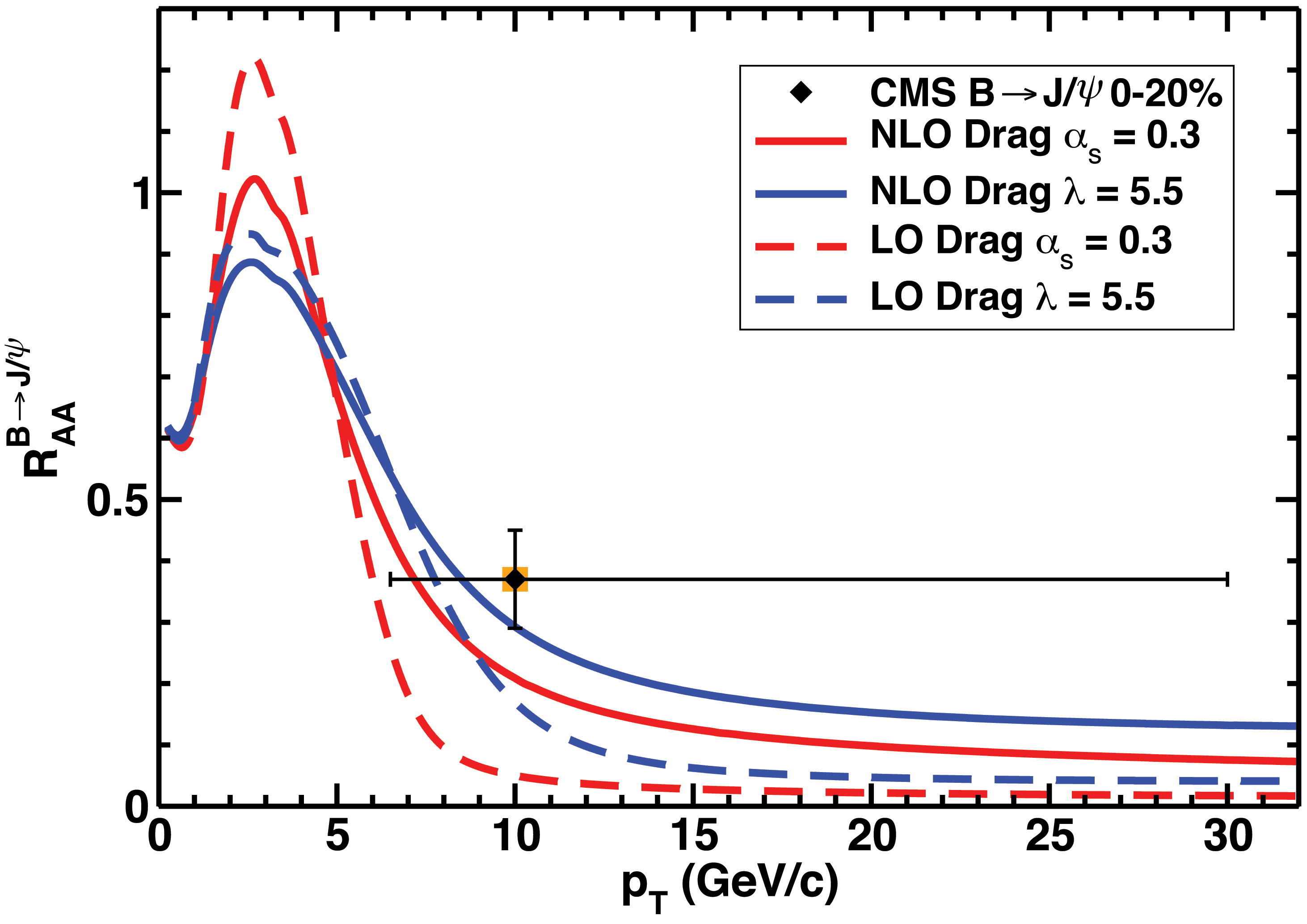}
		\label{subfig:LHCBJpsi}
	}\\[-.1in]
     \caption{\label{fig:suppression}(\protect\subref*{subfig:RHICe}) Open heavy flavor electron $R_{AA}(p_T)$ at RHIC measured by PHENIX \protect\cite{Adare:2010de} and STAR \protect\cite{Mustafa:2012jh} compared to predictions from heavy quark drag only (``LO'') and heavy quark drag with fluctuations (``NLO'') energy loss models.  (\protect\subref*{subfig:LHCD}) $D$ meson $R_{AA}(p_T)$ at LHC measured by ALICE \protect\cite{ALICE:2012ab} compared to the strongly-coupled energy loss predictions.  (\protect\subref*{subfig:LHCBJpsi}) Non-prompt $J/\psi$ meson $R_{AA}(p_T)$ at LHC measured by CMS \protect\cite{Chatrchyan:2012np} compared to the energy loss model predictions.  For curves labeled $\alpha_s \, = \, 0.3$, $\lambda \, = \, 4\pi\times0.3\times3$ and $T_{SYM} \, = \, T_{QCD}$; for curves labeled $\lambda \, = \, 5.5$, $T_{SYM} \, = \, T_{QCD}/3^{1/4}$.}
\end{figure}

\fig{suppression} shows the main results of this paper: predictions for $R_{AA}(p_T)$ for electrons from open heavy flavor at RHIC, $D$ meson suppression at LHC, and non-prompt $J/\psi$ suppression at LHC all compared to data \cite{Adare:2010de,Mustafa:2012jh,ALICE:2012ab,Chatrchyan:2012np}.  When including only the leading order drag without fluctuations, the model significantly oversuppresses the nuclear modification factor compared to data.  But when the correct fluctuations are included, the predictions---within their regime of applicability---are in qualitative agreement with all current suppression data.  

To expand on the point of the regime of applicability, while the electron prediction below $p_T \, \lesssim \, 3$ GeV/c is not in particularly good agreement with data, the electron production and/or fragmentation functions are untrustworthy below this momentum scale.  The regime of applicability is also crucial to note as it explains the dramatic increase in $R_{AA}$ for electrons and $D$ mesons for momenta beyond a few GeV/c ($\gtrsim \, 4$ GeV/c for electrons and $\gtrsim \, 10$ GeV/c for $D$ mesons): self-consistently, one sees that at momenta of order the speed limit, the momentum fluctuations begin to dominate at the momenta at which the setup of the problem is breaking down.  The lack of inclusion of Larmor-like energy loss \cite{Mikhailov:2003er,Fadafan:2008bq} due to what would be significant deceleration becomes a poor approximation, as is the white noise assumption for the fluctuations\footnote{Since the fluctuations increase dramatically with $\gamma$, a rare few of the quarks can actually unphysically runaway to arbitrarily large momenta; the quarks fluctuate between increasingly large positive and negative momenta for each time step.  In the code, there is a cutoff of 1000 GeV/c, after which the quark is considered unphysical.  The number of these ``lost'' quarks is $\mathcal{O}(10^{-4}\%)$.}.  When that important physics is included, the rapid rise in $R_{AA}(p_T)$ above $p_T \, \gtrsim \, 4-5$ GeV/c will be tamed, although the magnitude of the effect is not yet known or estimable.  

In previous work \cite{Horowitz:2010dm}, drag-only strong-coupling energy loss gave a good quantitative description of RHIC electron data, but, when using the same parameters, predicted an oversuppression of $D$ mesons at LHC \cite{Horowitz:2011wm}, due to the very sensitive $\mu\,\sim\,T^2$ nature of the drag.  In this calculation there is already an oversuppression at RHIC.  It is precisely this considerable sensitivity to temperature that leads to an oversuppression at RHIC in this calculation: the VISHNU hydrodynamics \cite{Shen:2011eg,Qiu:2011hf} is significantly hotter than the Glauber model used in the previous work \cite{Horowitz:2010dm}.  

\begin{figure}
	\includegraphics[width=\columnwidth]{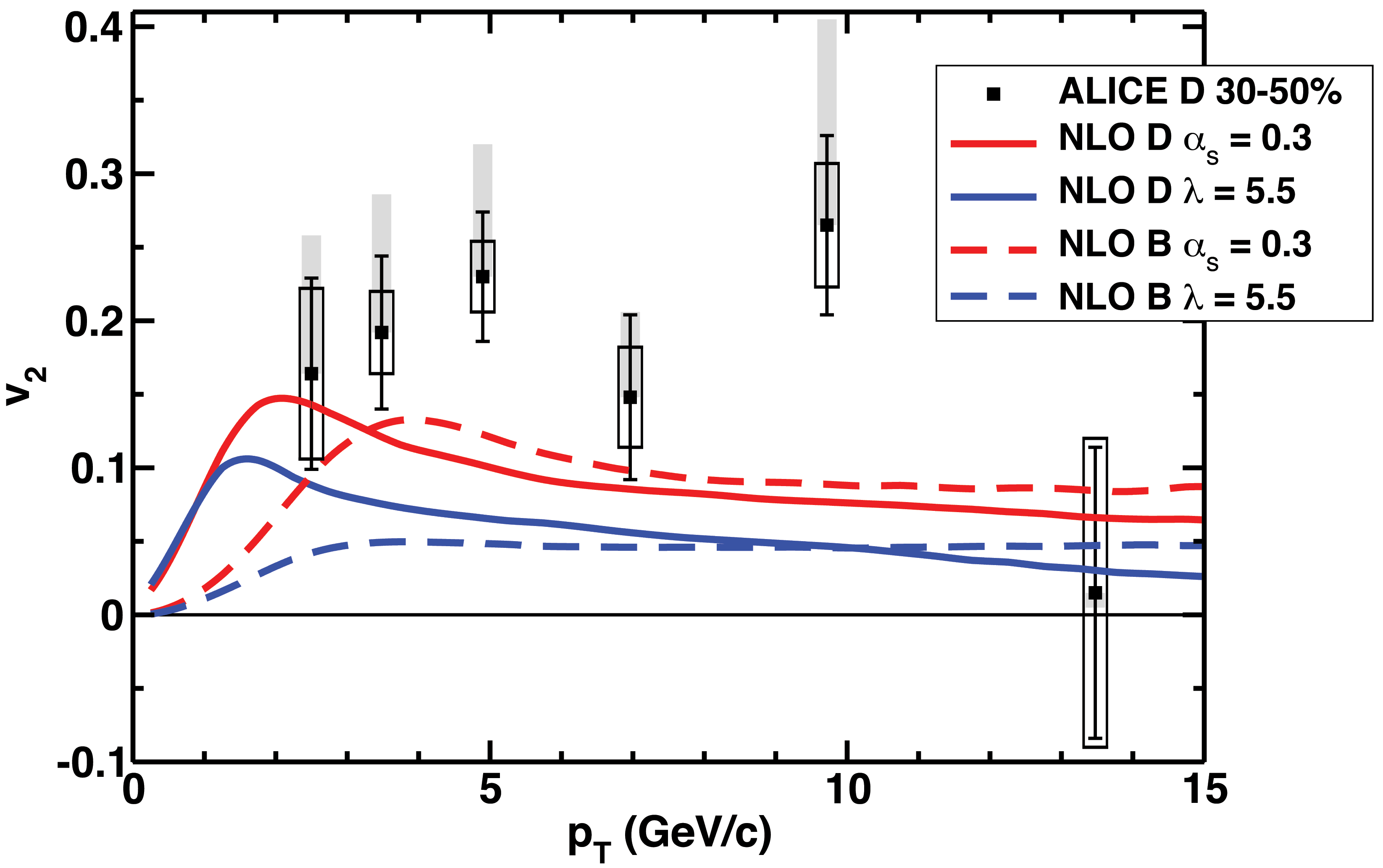}
	\caption{\label{fig:v2} $v_2(p_T)$ for $D$ mesons as measured by ALICE \protect\cite{Abelev:2013lca} at LHC compared to the predictions from the heavy quark energy loss model including fluctuations. For comparison the fluctuating energy loss model predictions for $B$ meson $v_2$ are also included. Curves labeled $\alpha_s \, = \, 0.3$ use $\lambda \, = \, 4\pi\times0.3\times3$ and $T_{SYM} \, = \, T_{QCD}$; curves labeled $\lambda \, = \, 5.5$ have $T_{SYM} \, = \, T_{QCD}/3^{1/4}$.}
\end{figure}

One may also compare to other observables, for instance the azimuthal anisotropy of the $D$ mesons, as shown in \fig{v2}.  The energy loss model including fluctuations qualitatively agrees with the experimental results from ALICE \cite{Abelev:2013lca}, although with a generic underprediction for the $v_2$ at intermediate-$p_T$ values $p_T \, \lesssim \, 10$ GeV/c.  This underprediction in $D$ mesons is reminiscent of the underprediction of $v_2$ for light flavor observables in pQCD-based energy loss models \cite{Shuryak:2001me,Horowitz:2005ja} and, like in the light flavor case, probably indicates a modification of the fragmentation process that enhances the anisotropy in this momentum range.  Future measurements with increased statistics will, as in the light flavor case \cite{ATLAS:2011ah,Horowitz:2011cv}, presumably show that the $D$ meson $v_2$ decreases as a function of $p_T$ to levels similar to the energy loss predictions.  Predictions for $B$ meson $v_2$ from the energy loss model are also shown in \fig{v2}; surprisingly, the model predicts $B$ meson $v_2$ of a similar size as the $D$ meson $v_2$.  It will be very interesting to see if measurements of the $B$ meson $v_2$ are of a similar magnitude to the $D$ meson $v_2$ as predicted here.

\begin{figure}
	\includegraphics[width=\columnwidth]{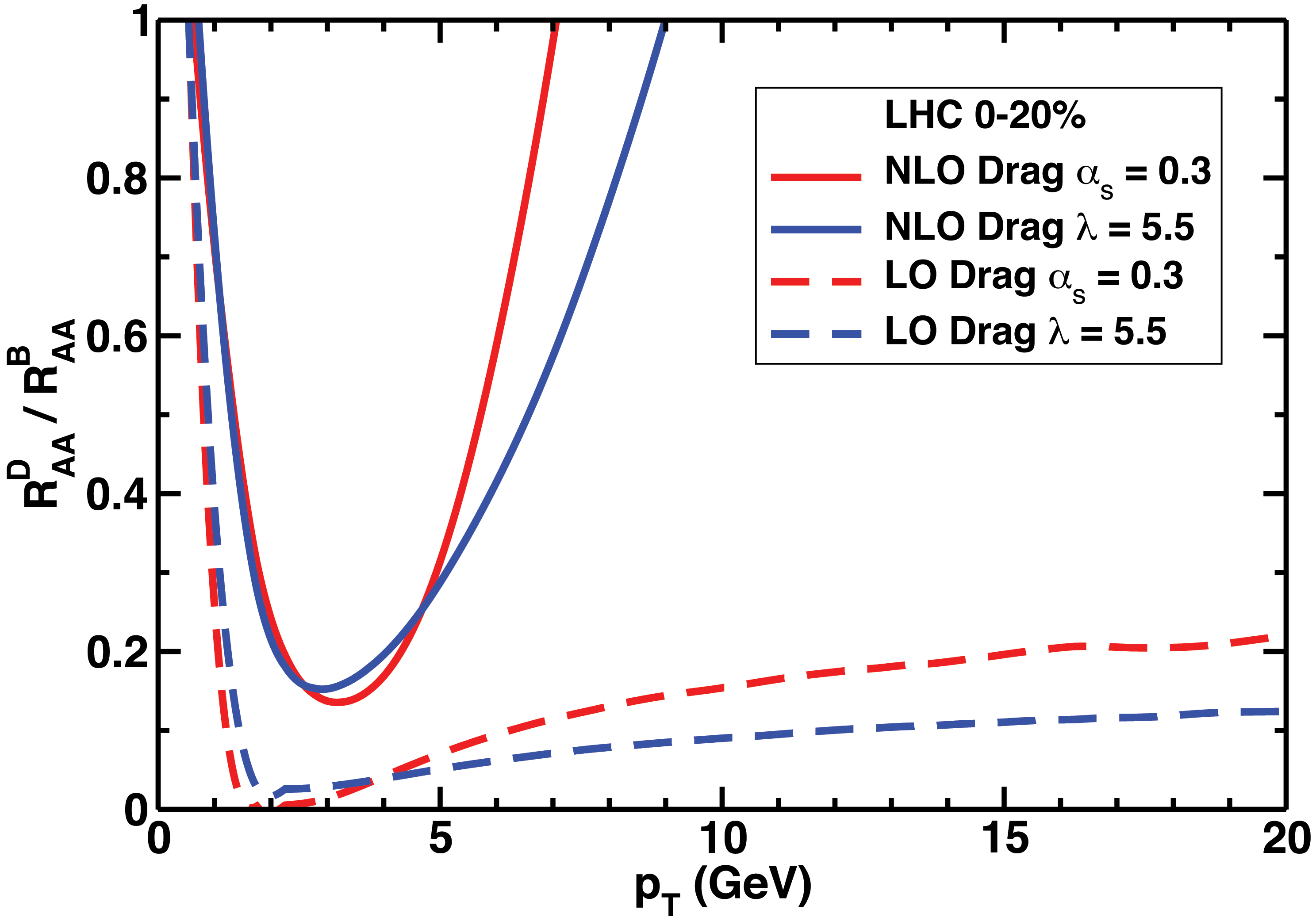}
	\caption{\label{fig:double} The ratio of $D$ meson $R_{AA}(p_T)$ to $B$ meson $R_{AA}(p_T)$ from the strongly-coupled heavy quark energy loss model at leading order (dashed) and including fluctuations (solid). Curves labeled $\alpha_s \, = \, 0.3$ use $\lambda \, = \, 4\pi\times0.3\times3$ and $T_{SYM} \, = \, T_{QCD}$; curves labeled $\lambda \, = \, 5.5$ have $T_{SYM} \, = \, T_{QCD}/3^{1/4}$.}
\end{figure}

Ideally, one would very much like some kind of smoking gun experimental measurement to distinguish between weakly-coupled energy loss mechanisms in the plasma compared to strongly-coupled energy loss mechanisms.  The double ratio of the nuclear modification factor of $D$ mesons to $B$ mesons was proposed as just such a distinguishing measurement \cite{Horowitz:2007su}.  It was shown in \cite{Horowitz:2007su} that the leading order drag prediction for the double ratio is flat as a function of $p_T$ and at a value of approximately $m_D/m_B \, \simeq \, 0.2$.  On the other hand, pQCD-based energy loss models predict an insensitivity to the mass of the heavy quark when $p_T \, \gg \, M_Q$, and thus the double ratio starts around 0.2 for $p_T \, \sim \, 10$ GeV/c and approaches 1 from below by $p_T \, \sim \, 100$ GeV/c.  It is natural to ask whether the strong-coupling prediction for the double ratio of $D$ to $B$ meson nuclear modification factors is affected by the inclusion of fluctuations.  It is shown in \fig{double} that fluctuations completely change the prediction for strong-coupling: the extremely rapid rise in $R_{AA}(p_T)$ for $D$ mesons due to fluctuations is reflected in an extremely rapid rise in the double ratio, with the double ratio actually going above 1 at $p_T \, \sim \, 7$ GeV/c.  Of course it is precisely at this momentum range that the drag calculation is breaking down; an improved calculation that includes the additional Larmor-like energy loss and autocorrelations in the fluctuations will bring down the double ratio prediction, although it is not yet clear by how much.  It seems likely that, for single particle observables, the best experimental measurement to distinguish between heavy quark drag-like energy loss and pQCD-like energy loss will come from $B$ meson $R_{AA}(p_T)$ alone: \fig{suppression} (\subref*{subfig:LHCBJpsi}) shows the drag (with fluctuations) prediction decreases significantly with $p_T$ and remains relatively constant while pQCD predicts that the $B$ meson $R_{AA}(p_T)$ will rise as a function of momenta for $p_T \, \gtrsim \, 25$ GeV/c.  One might notice that there is no prediction for $R_{AA}^{B\rightarrow J/\psi}(N_{part})$ given in this work.  Since the strong-coupling energy loss model qualitatively describes the most central nuclear modification factor for non-prompt $J/\psi$ mesons, and the model necessarily will predict that the nuclear modification factor goes to unity as collisions become more peripheral, it is difficult to imagine that the energy loss model will \emph{not} follow the experimentally measured trend \cite{CMSJpsinpart}.

\section{Conclusions and Outlook}
Hard probes, and, in particular, heavy quarks, provide a crucial test of our theoretical understanding of the physics of quark-gluon plasma.  Previous work \cite{Horowitz:2011wm} that included only the leading order energy loss \cite{Gubser:2006bz,Herzog:2006gh,CasalderreySolana:2006rq} for heavy quarks strongly-coupled to a strongly-coupled plasma as calculated from the AdS/CFT correspondence showed significant, falsified disagreement with data \cite{Adare:2006nq,ALICE:2012ab}.  The results presented in this paper in Figs.\ \ref{fig:suppression} and \ref{fig:v2}, for which energy loss fluctuations \cite{Gubser:2006nz} are correctly included for the first time in a realistic strong-coupling energy loss model for heavy quarks, are qualitatively consistent with data from RHIC \cite{Adare:2010de,Mustafa:2012jh} to LHC \cite{ALICE:2012ab,Chatrchyan:2012np,Abelev:2013lca} within the regime of applicability of the energy loss formulae, Eqs.\ (\ref{eq:langevin}) -- (\ref{eq:kappaL}).  

Along with the recent improvement in the treatment of light flavor energy loss in AdS/CFT \cite{Morad:2014xla}, in which very good agreement between the suppression of jets from strong-coupling and the preliminary CMS data \cite{CMS:2012rba} was shown, strong-coupling can now explain within a self-consistent picture of quark-gluon plasma dynamics a wide and impressive array of observables from low-momentum to high-momentum that includes the rapid applicability of hydrodynamics \cite{Chesler:2010bi}, the very small shear viscosity-to-entropy ratio \cite{Gubser:1996de,Kovtun:2004de}, the light flavor jet suppression \cite{Morad:2014xla}, and the suppression of decay products of heavy quarks (this work).  

From the specific perspective of heavy flavor energy loss from AdS/CFT, the next steps include further testing theoretical predictions against data---especially in comparison with predictions from pQCD---and in improving the theory itself.

Surprisingly, as shown in \fig{double}, the inclusion of fluctuations completely changes the leading order \cite{Horowitz:2007su} qualitative prediction of the double ratio of $D$ to $B$ nuclear modification factors from the strong-coupling energy loss model: with fluctuations, the double ratio rises rapidly as a function of meson momentum, broaching unity at a momentum scale somewhat above where the calculation is inapplicable due to the speed limit of the setup, \eq{speedlimit}.  Although the rise is now more rapid than that predicted by pQCD \cite{Horowitz:2007su}, it is not known how much the momentum rise in the AdS/CFT prediction will be softened as necessary corrections are included to extend the regime of applicability of the calculation to $p_T$ above a handful of GeV/c; it is therefore difficult at this stage to use an experimental measurement of the double ratio to test strong- versus weak-coupling physics.  

The anisotropy of heavy flavor decay products provides a strong- versus weak-coupling energy loss comparison, although of limited utility.  From strong-coupling, as shown in \fig{v2}, $B$ meson $v_2(p_T)$ is predicted to be of a similar magnitude to $D$ meson $v_2(p_T)$.  Due to mass effects one expects pQCD to predict a smaller anisotropy for $B$ mesons than for $D$ mesons, but it does not seem that predictions for $v_2^B(p_T)$ are extant in the literature.  This prediction probably has limited use for distinguishing between strong- and weak-coupling dynamics as a precision differentiation of $v_2$ for $D$ versus $B$ mesons poses a significant experimental challenge.  Additionally, it is not clear the extent to which non-perturbative hadronization effects differ for $D$ versus $B$ meson production in heavy ion collisions at relatively low $\lesssim \, 10$ GeV/c momentum at which the AdS/CFT calculations are within their regime of applicability for $D$ mesons.  

A likely much more useful prediction comes from the nuclear modification factor of $B$ mesons.  As can be seen in \fig{suppression} (\subref*{subfig:LHCBJpsi}), the strong-coupling energy loss model predicts a large $B$ meson suppression, $R_{AA}(p_T) \, \sim \, 0.1$, out to $p_T \, \sim \, 30$ GeV/c; in contrast, pQCD predicts a much smaller $B$ meson suppression with $R_{AA}(p_T) \, \sim \, 0.5$ \cite{Horowitz:2011wm}.  It is worth emphasizing that the strong-coupling energy loss derivations and implementations are under increasing theoretical control as the mass of the quark increases, so future measurements related to $b$ quarks will be especially interesting to see.

Since momentum fluctuations play such an important role in strongly-coupled heavy quark energy loss, and these fluctuations grow characteristically with momentum, a comparison between theory and data related to correlations of pairs of heavy quarks would likely be very enlightening.  The comparison would be most meaningful for the decay products of $b$ quarks at high momentum, $10 \, \lesssim \, p_T \, \lesssim \, 100$ GeV/c: above $\sim\,10$ GeV/c any effect that leads to the enhanced $v_2$ at intermediate-$p_T$ should be small, and one should be able to trust the theoretical calculation below $\sim\,100$ GeV/c.  The determination of these correlations from this energy loss model is left for future work. 

Improvements to the theoretical calculations that need to be done would increase the regime of applicability of the energy loss formulae and model, thus allowing for a meaningful comparison to more data.  The heavy flavor electron predictions could be extended to lower momenta if their production in $p+p$ collisions is brought under better phenomenological control.  Both the heavy flavor electron and $D$ meson predictions could be extended to higher momenta if the theoretical and numerical complications related to the speed limit, \eq{speedlimit}, could be overcome.  It is likely that an entirely different approach to the energy loss calculation, both analytic and numerical, will be required: since the setup of a heavy quark moving at a constant speed is breaking down, presumably one must allow the heavy quark to slow down, which, in turn, will likely mean that full numerical solutions to the worldsheet evolution including thermal fluctuations will be needed.  This important, non-trivial research is also left for future work.

\acknowledgments
The author wishes to thank the South African National Research Foundation and SA-CERN for support and CERN for hospitality.  The author also wishes to thank Derek Teaney and Urs Wiedemann for useful discussions.

\appendix

\section{The J\"uttner Distribution}
\label{juttner}
The number of particles $N$ of a relativistic gas in the rest frame of the gas in an $n$-spatial-dimensional box is given by the integral over the invariant phase space of the J\"uttner distribution \cite{He:2013zua,Csernai:1994xw,PhysRevLett.99.170601}, the relativistic generalization of the Maxwell-Boltzmann distribution,
\begin{equation}
	N = \int d^nx \, d^np \, A\, e^{-p^0/T},
\end{equation}
in units of $\hbar \, = \, c \, = \, k \, = 1$, where $p^0 \, = \, \surd(\vec{p}^2+m^2)$ is the relativistic energy of a particle of mass $m$, which can be 0.  $T$ is the well-defined temperature of the gas in its rest frame, and $A$ is a normalization constant.

Since we are interested in comparing to a numerical simulation with fixed number of particles $N$ we may readily determine $A$:
\begin{align}
	\label{eq:n}
	N & = A\,V\,\Omega_{n-1}\int_0^\infinity p^{n-1}dp\,e^{-\sqrt{p^2+m^2}/T},
\end{align}
where
\begin{align}
	\Omega_d = \frac{2\pi^{\frac{d+1}{2}}}{\Gamma\big(\frac{d+1}{2}\big)}
\end{align}
is the volume of a $d$-dimensional sphere; e.g., $\Omega_2 \, = \, 4\pi$.  The integral in \eq{n} may be computed analytically in terms of Gamma and modified Bessel functions, such that
\begin{equation}
	A = \frac{1}{2}\frac{N}{V}\Big( \frac{m}{2\pi T} \Big)^{\frac{n-1}{2}}\frac{1}{m^n K_{\frac{n+1}{2}}(m/T)}.
\end{equation}

Thus in the rest frame of the plasma the momentum distribution of the $N$ particles is given by
\begin{align}
	\frac{dN}{d^np} = \frac{1}{2}N\Big( \frac{m}{2\pi T} \Big)^{\frac{n-1}{2}} \frac{e^{-p^0/T}}{m^n K_{\frac{n+1}{2}}(m/T)}.
\end{align}
Picking out one special direction, say the $z$-direction, one may find the $p^z$-differential distribution of particles in the rest frame of the plasma:
\begin{multline}
	\frac{dN}{dp^z}
	= \frac{1}{2}N\Big( \frac{m}{2\pi T} \Big)^{\frac{n-1}{2}} \frac{1}{m^n K_{\frac{n+1}{2}}(m/T)} \\ \times \int d^{n-1}p_T\,e^{-\sqrt{p_T^2+(p^z)^2+m^2}/T}.
\end{multline}
If one is interested in the distribution of particles in a frame that is boosted along, say, the $z$-direction at a speed $\beta$ compared to the rest frame of the plasma, then one finds that in the primed coordinate system
\begin{widetext}
 \begin{align}
  	\frac{dN}{dp^{z'}}
  	& = \frac{1}{2}N\Big( \frac{m}{2\pi T} \Big)^{\frac{n-1}{2}} \frac{1}{m^n K_{\frac{n+1}{2}}(m/T)} \int d^{n-1}p_T\,\frac{dp^z}{dp^{z'}}e^{-\sqrt{p_T^2+\big(p^z(p^{z'})\big)^2+m^2}/T} \\
 	& = \frac{1}{2}N\Big( \frac{m}{2\pi T} \Big)^{\frac{n-1}{2}} \frac{1}{m^n K_{\frac{n+1}{2}}(m/T)} \Omega_{n-2} \nonumber\\
 	& \qquad\times\int p_T^{n-2}dp_T \, \bigg(\frac{\gamma}{\sqrt{p_T^2+(p^{z'})^2+m^2}} \big[ \sqrt{p_T^2+(p^{z'})^2+m^2}-\beta\,p^{z'} \big]\bigg) e^{-\sqrt{p_T^2 + \gamma^2\big( p^{z'}-\beta\sqrt{p_T^2+(p^{z'})^2+m^2} \big)^2 + m^2}/T} \\
 	& = \frac{N\,\gamma^2}{\sqrt{2\pi m T}}\Big( \frac{m^2+(p^{z'})^2}{\gamma^2 \, m^2} \Big)^{n/4} \frac{e^{\gamma\,\beta\,p^{z'}/T}}{K_{\frac{n+1}{2}}(m/T)}\Big( K_{\frac{n}{2}}\big( \gamma\sqrt{(p^{z'})^2+m^2}/T \big) - \frac{\beta\,p^{z'}}{\sqrt{(p^{z'})^2+m^2}}K_{\frac{n-2}{2}}\big( \gamma\sqrt{(p^{z'})^2+m^2}/T \big)\Big).
 \end{align}

\section{Semi-leptonic and \texorpdfstring{$J/\psi$}{J/psi} Fragmentation Functions}
\label{ffs}
The spectrum of electrons $dN^e/dp_T^e(p_T^e)$ (or $J/\psi$ mesons) produced in the lab frame by a spectrum of heavy mesons (in this case a $D$ meson for specificity) that decays into an electron of energy $E$ with probability $P(E)$ in the rest frame of the heavy flavor meson is given by:
\begin{align}
	\label{eq:electrons}
	\frac{dN^e}{dp_T^e}(p_T^e) & = \int dp_T^D \frac{d\Omega}{4\pi} dE \frac{dN^D}{dp_T^D}(p_T^D) P(E) \delta\Big(p_T^e - E\sqrt{\gamma_D^2\big( \sin(\theta)\cos(\phi)+\beta_D \big) + \sin^2(\theta)\sin^2(\phi)}\Big) \\
	\label{eq:elecuse}
	& = \int_{p^D_{min}}^\infinity dp_T^D \int_{\theta_{min}}^{\theta_{max}} \sin(\theta) d\theta \int_{\phi_{min}}^{\phi_{max}} \frac{d\phi}{\pi} J(\theta,\,\phi) \frac{dN^D}{dp_T^D}(p_T^D) P\big(p_T^e \, J(\theta,\,\phi)\big) \theta\big(|\sec(\theta)|J^{-1}(\theta,\,\phi) - \mathrm{csch}(y_{max})\big),
\end{align}
where
\begin{align}
	J(\theta,\,\phi) = \left[ \gamma_D^2 \big( \sin(\theta)\cos(\phi)+\beta_D \big)^2 + \sin^2(\theta)\sin^2(\phi) \right]^{-1/2},
\end{align}
\newpage
\end{widetext}

\begin{align}
	p^D_{min} & = \max\big(0,\,m_D \frac{(p_T^e)^2 - E_{max}^2}{2 \, p_T^e \, E_{max}}\big) \nonumber\\
	\theta_{min} & = \max\Big( 0,\, \Re\big(\theta_{crit}^+(E_{max})\big),\, \Re\big(\theta_{crit}^-(E_{min})\big) \Big) \nonumber\\
	\theta_{max} & = \min\Big( \pi/2,\, \Re\big(\sin^{-1}(\beta_D + p_T^e / \gamma_D \,E_{min})\big) \Big) \nonumber\\
	\phi_{min} & = \Re\big(\phi_{crit}(E_{min})\big) \nonumber\\
	\phi_{max} & = \Re\big(\phi_{crit}(E_{max})\big) \\
	\theta_{crit}^\pm(E) & = \sin^{-1}( \mp \beta_D \pm p_T^e / \gamma_D E) \nonumber\\
	\phi_{crit}(E) & = \sin^{-1}\left( \frac{1}{\beta_D\sin(\theta)}\Big( -1 \right. \nonumber\\
	&\qquad + \left.\sqrt{1 - \beta_D^2 - \big( \frac{\sin(\theta)}{\gamma_D} \big)^2 + \big( \frac{p_T^e}{\gamma_D\,E} \big)^2} \Big) \right).\nonumber
\end{align}
$E_{max}$ is the largest energy electron and $E_{min} \, = \, m_e$ is the lowest energy electron that can be emitted in a semi-leptonic decay in the rest frame of the heavy meson, and $y_{max}$ is the largest rapidity at which electron measurements are reported (there is an assumption that measurements are restricted symmetrically about $y \, = \, 0$).  Note that the $\theta$ in the second line of \eq{electrons} is the usual Heaviside step function, and
\begin{align}
	\gamma_D & = \frac{\sqrt{p^2+m_D^2}}{m_D} \\
	\beta_D & = \frac{p}{\sqrt{p^2+m_D^2}}.
\end{align}
The explicit probability functions that were used in \eq{electrons} are (with some numbers truncated slightly for visibility, although it should not affect results)
\begin{widetext}
\begin{align}
\label{eq:pofd}
	P^{D\rightarrow e}(E) & = (E - E_{min}^D)(E_{max}^D - 
  E) (23.3123 E + 135.0847 E^2 - 685.7559 E^3 
  + 1206.6398 E^4 \nonumber \\
  & \qquad - 
   1013.2945 E^5 
   + 335.7252 E^6)/1.0015 \\
   P^{B\rightarrow e}(E) & = (E - E_{min}^B)(E_{max}^B - 
  E) (0.0015939 - 0.002299 E + 
   0.069117 E^2 - 0.089862 E^3 
   \nonumber\\ & \qquad + 
   0.054773 E^4 - 0.012189 E^5)/
 0.050435 \\
 P^{B\rightarrow D\rightarrow e}(E) & = -0.003556 + 
E(-0.359519 + 
   E(27.194146 + 
      E(-119.476261  + 
         E(244.65394 \nonumber\\
                  & \qquad + 
            E(-292.041775 + 
               E(218.040963 + 
                  E(-103.278527 +
                   E(30.170577 \nonumber\\
                  & \qquad + 
                   E(-4.960071 + E(0.351141)))))))))) 
                   (2.65 - 
   E)/1.004464 \\
\label{eq:pofjpsi}
  P^{B\rightarrow J/\psi}(E) & = (E (\sqrt{E^2 - m_{J/\psi}^2} - 2) (126.32439 (E^2 - m_{J/\psi}^2)^2 + 
    21.848881 (E^2 - m_{J/\psi}^2)^3 -  \nonumber\\
    & \qquad
    1.728567 \sqrt{E^2 - m_{J/\psi}^2} + 
    1.721525 (E^2-m_{J/\psi}^2) - 
    64.938832 (E^2 - m_{J/\psi}^2)^{3/2} \nonumber\\
    & \qquad - 
    90.585013 (E^2 - m_{J/\psi}^2)^{5/2}))\Big/\big(\sqrt{E^2 - m_{J/\psi}^2}
   E_m^2 (1.728567 - 1.723872 E_m + \nonumber\\
   & \qquad 
    32.899797 E_m^2 - 63.517521 E_m^3 + 
    51.249069 E_m^4 - 19.183254 E_m^5 + 
    2.731110 E_m^6)\big),
\end{align}
\end{widetext}
where $E_m \, = \, (m_B^2 - m_{J/\psi}^2)/2m_B$. See \tab{Plimits} for the maximum and minimum allowed energies for each of the above $P$ functions. 

\begin{table}
	\centering
	\begin{tabular}{|c|c|c|c|c|}
		\hline
		& $D\rightarrow e$ & $B\rightarrow e$ & $B\rightarrow D\rightarrow e$ & $B\rightarrow J/\psi$ \\
		\hline
		$E_{min}$ & $m_e$ & $m_e$ & $m_e$ & $m_{j/\psi}$ \\
		\hline
		$E_{max}$ & 0.95 GeV & 2.3 GeV & 2.6 GeV & $\sqrt{E_m^2 + m_{J/\psi}^2}$ \\
		\hline
	\end{tabular}
	\caption{\label{tab:Plimits} Maximum and minimum allowed energies for the semi-leptonic electron decay fragments of $D$ and $B$ mesons and the $J/\psi$ mesons from $B$ mesons in the rest frame of the decaying heavy meson. $m_e \, = \, 0.0005$ GeV and $m_{J/\psi} \, = \, 3.096916$ GeV.}
\end{table}

%******************************references***************************

\bibliography{refs}

\end{document}